\documentclass[superscriptaddress,aps,prd,twocolumn,showpacs,nofootinbib]{revtex4-1}
\usepackage{etex}
\usepackage{amsmath,amssymb,amsthm}
\usepackage{ragged2e}
\usepackage[colorlinks=true,citecolor=blue,urlcolor=blue]{hyperref}
\usepackage[pdftex]{graphicx}  
\usepackage{verbatim}   
\usepackage{color}      
\usepackage{subfigure}  
\usepackage{hyperref}   
\usepackage{mathrsfs}
\usepackage{booktabs}
\usepackage{wrapfig}
\usepackage{epstopdf}
\begin{document}
\title{Effect of inhomogeneities on the propagation of gravitational waves from binaries of compact objects}
\author{Shashank Shekhar Pandey}   \email{shashankpandey7347@gmail.com}
\affiliation{Department of Astrophysics and Cosmology, 
S. N. Bose National Centre for Basic Sciences, JD Block, Sector III, 
Salt lake city, Kolkata-700106, India}
\author{Arnab Sarkar}         \email{arnabsarkar@bose.res.in, arnab.sarkar14@gmail.com}
\affiliation{Department of Astrophysics and Cosmology, 
S. N. Bose National Centre for Basic Sciences, JD Block, Sector III, 
Salt lake city, Kolkata-700106, India}
\author{Amna Ali }  \email{amnaalig@gmail.com}
\affiliation{Department of Mathematics, Jadavpur University, Kolkata-700032, India}
\author{A. S. Majumdar }   \email{archan@bose.res.in}
\affiliation{Department of Astrophysics and Cosmology, 
S. N. Bose National Centre for Basic Sciences, JD Block, Sector III, 
Salt lake city, Kolkata-700106, India}
\date{\today}
\begin{abstract}
\begin{center}
\textbf{Abstract}
\end{center}
\begin{small}
We consider the propagation of gravitational waves in the late time Universe with the presence of structure. Before detection, gravitational waves emitted from distant sources have to traverse through regions of spacetime which are far from smooth and homogeneous. We investigate the effect of inhomogeneities on the observables associated with the gravitational wave sources. In particular, we evaluate the impact of inhomogeneities on gravitational wave propagation by employing Buchert's framework of averaging. In context of a toy model within the above framework, it is first shown how the redshift versus distance relation gets affected through the averaging process. We then study the variation of the redshift dependent part of the observed gravitational wave amplitude for different combination of our model parameters. We show that  the variation of the gravitational wave amplitude with respect to redshift can deviate significantly compared to that in the $\Lambda$CDM-model. Our result signifies the importance of local inhomogeneities on precision measurements of parameters of gravitational wave sources.

\end{small}
\end{abstract}
\maketitle


\begin{small}
\vspace{0.5cm}
\section{Introduction }

The concept of Gravitational waves exists in Einstein's theory of General Relativity since 1916 \cite{Einstein1, Einstein2}. These are ripples in spacetime, caused by any mass (or energy) in a motion such that the concerned second mass-moment (or quadrupole moment in transverse-traceless gauge) has non-vanishing second-order time-derivative. The recent excitement in the field  stems from several detections of gravitational waves from compact binary mergers since the first report by the LIGO and VIRGO scientific collaborations \cite{Abbott_et_al, Ligo-virgo2, Ligo-virgo3, Ligo-virgo4, Ligo-virgo5, Ligo-virgo6}. The observed parameters of gravitational waves are significant for inferring parameters associated with the sources, such as their mass range and merger rates. Generally, for the data analysis of gravitational wave observations, it is considered that the waves propagate from the source to the detector through a homogeneous and isotropic FLRW-Universe.

\par The standard $\Lambda$CDM model of cosmology describes the Universe as an FLRW spacetime, which is statistically homogeneous and isotropic at very large length scales. However, this homogeneity does not hold at smaller scales. Various tests from cosmological observations like Sloan Digital Sky Survey \cite{Sloan} and the WiggleZ Dark Energy Survey \cite{WiggleZDE} indicate a transition from homogeneity to inhomogeneity at smaller scales. Current estimates to analyze large scale fluctuations in the luminous red galaxy samples
based on higher-order correlations have found significant (more than 3 $\bar{\sigma}$) deviations from the $\Lambda$CDM mock catalogues on samples as large as 500 $h^{-1}$ Mpc \cite{Weigand_scale}. Thus, inhomogeneities due to structures may have important effects on length scales even as large as
500 $h^{-1}$ Mpc. It may be also pertinent to mention here that even if the distribution can be regarded as homogeneous, it does not necessarily mean that the model belongs to the FLRW family. The present observations of gravitational waves \cite{Abbott_et_al, Ligo-virgo2, Ligo-virgo3, Ligo-virgo4, Ligo-virgo5, Ligo-virgo6} pertain to sources which lie well within the scale at which there exists overall homogeneity in the Universe. Thereby, the investigation of the effect of inhomogeneities on propagation of gravitational wave may be of significance for precision measurements in the emerging  field of gravitational wave astronomy. 

\par In order to investigate the effects of local inhomogeneities on the dynamics at larger scales, an averaging procedure is necessary. This effect is generally termed `backreaction'. The concept of averaging was first introduced in general relativity in 1963 \cite{Shirkov_et_al}, but the proposed procedure was not covariant. A concrete description of the backreaction concept in this regard was given by Ellis  \cite{Ellis}. Further, the gravitational correlation by employing metric perturbations was also studied \cite{Futamase}. Futamase and Sasaki analyzed the effect of these inhomogeneities on the propagation of electromagnetic waves \cite{Futamase_Sasaki} using an approximate metric for describing an inhomogeneous Universe. Using these, they derived a general distance-redshift relation for the inhomogeneous Universe. A covariant averaging scheme of tensors via bilocal operators was provided, leading to an averaged version of Einstein's equations \cite{Zalaletdinov-1, Zalaletdinov-2}. On the other hand, a formalism restricted to averaged scalars was proposed by Buchert \cite{Buchert-1, Buchert-2}, which has led to several exciting applications of backreaction due to inhomogeneities
\cite{Schwarz, Rasanen, Wiltshire, Kolb_et_al, Ishibashi_et_al, Coley, Rasanen-2, Rasanen-3, Gasperini_et_al, Gasperini_et_al-2, bose, Fleury_et_al, bose2, Fleury_et_al-2, Fleury_et_al-3, Bagheri_et_al, AmnaAli, Koksbang}.

\par In Buchert's formalism \cite{Buchert-1, Buchert-2, Wiegand_et_al} the time evolution of spatial averages are not identical to those computed using the evolved spatial metric. These differences introduce the backreaction term, which is geometrical in nature and quantifies deviations from the usual behaviour of the Friedmann equations. Buchert's averaging technique provides a unique way of explaining the large scale expansion of space without oversimplifying the Universe by considering it homogeneous. Several works have been done attempting to explain the accelerated expansion of the late-time Universe through backreaction \cite{Schwarz, Rasanen, Wiltshire, Kolb_et_al, Ishibashi_et_al}. The cosmological backreaction requires special attention in the present era of precision cosmology.

\par Consideration of the effect of local inhomogeneities leads to certain interesting effects on the propagation of electromagnetic waves in averaged Universe. Studies investigating the motion of photons in an averaged geometry have been performed \cite{Coley, Rasanen-2, Rasanen-3}, with some employing a gauge-invariant approach for averaging on the past null cone \cite{Gasperini_et_al, Gasperini_et_al-2}. Other works have explored light propagation in inhomogeneous Swiss-Cheese models by simulating the Hubble-diagram \cite{Fleury_et_al, Fleury_et_al-2, Fleury_et_al-3}. An averaged version of the null geodesic equation has been derived \cite{Bagheri_et_al}  showing that the light propagation equation contains an effective Hubble parameter for averaged spacetime, which is different from that of the homogeneous spacetime.
Modified cosmological-distance versus redshift relations ensue in the context of averaging over inhomogeneities \cite{Koksbang,koksnew1, heinesen}. This modification has certain interesting implications for detecting signals of inhomogeneity through probes of the cosmic expansion rate \cite{koksnew2}.

\par Gravitational waves act as complementary messengers to electromagnetic waves opening up a new window to the physics of the  Universe. Both electromagnetic and gravitational waves can be used together to study the expansion history of the Universe, which is essential for the understanding of current cosmic acceleration and the nature of gravity itself. Further, gravitational waves have special significance for those sources which do not emit any electromagnetic signals. The scientific insights unfolding from gravitational wave detection are required to further our understanding of multiple domains of physics, astrophysics and cosmology. In this work, our goal is to study the effect of local inhomogeneities in the Universe on the propagation of gravitational waves. 
 
\par The formalism of this work is based on  Buchert's backreaction scheme \cite{Buchert-1, Buchert-2, Wiegand_et_al}. In the context of Buchert's averaging procedure, we employ a simplistic model of a two-partitioned Universe {\it viz.}, all inhomogeneities are clubbed into overdense (wall) and underdense (void) regions. Our motivation is to investigate the change in observed amplitude of the gravitational wave signal due to averaging over various combinations of the fractions of these two types of partitions in comparison to the case where the gravitational wave is assumed to traverse through a completely homogeneous and isotropic spacetime described by the $ \Lambda$CDM model. In the above framework, we consider gravitational waves from binaries of compact objects like black holes and neutron stars in their early inspiral stage. Our analysis demonstrates that there could be a significant deviation in the gravitational wave amplitude as an effect of backreaction due to inhomogeneities present in the intervening spacetime between the source and detector.

\par The paper is organized as follows. A brief description of our two-partitioned model and ansatz in context of the Buchert's backreaction formalism is presented in section \ref{Sec2}. In section \ref{Sec3}, the modification of the redshift-distance relation due to the averaging procedure is presented. The modification of the redshift dependent part of gravitational wave amplitude from a binary of compact objects in our model is analyzed and compared to the case in the $\Lambda $CDM model in section \ref{Sec4}. Finally, we present some concluding remarks in section \ref{Sec5}. 
    
\section{A model of backreaction in Buchert's formalism}\label{Sec2}
In Buchert's formalism \cite{Buchert-1, Buchert-2, Wiegand_et_al} the Einstein's equations are decomposed into a set of dynamical equations for scalar quantities. We can take the spacetime to be foliated into flow-orthogonal hypersurfaces with an inhomogeneous 3-metric $ g_{ij}$ and the line-element 
\begin{equation} \label{1.1}
ds^{2}= -dt^{2} + g_{ij} dx^{i} dx^{j} \, ,  
\end{equation}
where $t$ is the proper time of the hypersurfaces and $x^{i} $ are the spatial coordinates in the hypersurfaces of constant $t$. The spatial average of any scalar quantity $f $ is defined on the constant-time hypersurfaces as 
\begin{equation} \label{1.2}
\langle f \rangle (t, \vec{x}) = \frac{\int d^{3}x \sqrt{g(t, \vec{x})} f(t, \vec{x}) }{\int d^{3}x \sqrt{g(t, \vec{x})} }   \, , 
\end{equation}
where $\vec{x}$ denotes the position 3-vector. The averaged volume scale-factor of a compact domain $\mathcal{ D} $ is defined as 
\begin{equation} \label{1.3}
a_{\mathcal{D}} = \Big( \frac{\int d^{3}x \sqrt{g(t, \vec{x})}}{\int d^{3}x \sqrt{g(t_{0}, \vec{x}) }}  \Big)^{1/3} \,  , 
\end{equation} 
where $t_{0} $ is any reference-time, which is generally chosen to be the present time. The spatial averaging and the time-evolution of any scalar quantity does not commute in this formalism, i.e.,
\begin{equation} \label{1.4}
\partial_{t} \langle f \rangle_{\mathcal{D}} -  \langle \partial_{t } f \rangle_{\mathcal{D}} = \langle f \theta \rangle_{\mathcal{D}} - \langle f \rangle_{\mathcal{D}} \langle\theta  \rangle_{\mathcal{D}} \, ,    
\end{equation}
where the quantity $ \theta $ denotes the time-rate of expansion of the spatial 3-metric: 
\begin{equation} \label{1.5}
\theta = (\sqrt{g(t , \vec{x})} )^{-1} \partial_{t} \sqrt {g(t , \vec{x}) }  \, . 
\end{equation}
In this approach the averaged equations for volume expansion and volume acceleration for the domain $\mathcal{D}$, with cosmological constant $\Lambda $ and the cosmic-fluid as an irrotational dust with density $ \rho $, are found by averaging the Hamiltonian, Raychaudhuri and energy-conservation equations, and are given by respectively, 
\begin{eqnarray} \label{1.6}
3 \frac{\ddot{ a}_{\mathcal{D}}}{a_{\mathcal{D}}} = -\frac{4\pi G }{c^{4}} \langle \rho \rangle_{\mathcal{D}} + \mathcal{Q}_{\mathcal{D}} + \Lambda  \,  ,   \\
3H_{\mathcal{D}}^{2} = \frac{8 \pi G}{c^{4}} \langle \rho \rangle_{\mathcal{D}} - \frac{1}{2} \langle \mathcal{R} \rangle_{\mathcal{D}} - \frac{1}{2}  \mathcal{Q}_{\mathcal{D}}  + \Lambda   \, ,  \\
\partial_t\langle\rho\rangle_D + 3H_D\langle\rho\rangle_D = 0  , 
\end{eqnarray}
where $\langle \mathcal{R} \rangle_{\mathcal{D}}$ is the averaged 3-Ricci scalar-curvature,
$H_{\mathcal{D}} $ is the averaged Hubble-parameter, $\Lambda$ is the cosmological constant which has been taken as zero here for our model and $\mathcal{Q}_{\mathcal{D}}$ is the backreaction term which quantifies the averaged effect of the inhomogeneities in the domain $\mathcal{D} $, defined as 
\begin{equation} \label{1.7}
\mathcal{Q}_{\mathcal{D}} = \frac{2}{3} ( \langle \theta^{2} \rangle_{\mathcal{D}} - \langle \theta \rangle^{2}_{\mathcal{D}}) - 2  \langle \sigma^{2} \rangle_{\mathcal{D}}  \,  , 
\end{equation}
where $\sigma^{2} = \frac{1}{2} \sigma_{ij} \sigma^{ij}$ is the shear-scalar of the cosmic-fluid. 
$\mathcal{Q}_{\mathcal{D}} $ and $\mathcal{R}$ are inter-related by the equation  
\begin{equation} \label{1.8}
\frac{1}{a_{\mathcal{D}}^{6}} \partial_{t} (a_{\mathcal{D}}^{6} \mathcal{Q}_{\mathcal{D}}) +  \frac{1}{a_{\mathcal{D}}^{2}} \partial_{t} (a_{\mathcal{D}}^{2} \langle \mathcal{R} \rangle_{\mathcal{D}} ) = 0 \, . 
\end{equation}  
The above equation (\ref{1.8}) signifies the fact that due to the inhomogeneities, the temporal-evolution of the average 3-Ricci scalar-curvature is different from that in the case of a homogeneous FLRW-Universe.


In the above framework we consider a simplified scenario where the domain $ \mathcal{D}$, which contains the path traversed by the gravitational wave from the source to observer, can be broadly classified into two types of regions or sub-domains 
\cite{Rasanen-2, bose, bose2, Koksbang, AmnaAli}: (i) the overdense region or `Wall' and (ii) the underdense region or `Void'. Note that in the present work we will not consider any possible attenuation of the gravitational wave amplitude due to shear effects.

\par  We aim to investigate the difference in the observed amplitude of gravitational waves in the two different cases, {\it viz}. (a) when the gravitational wave is assumed to propagate through a homogeneous and isotropic spacetime, described by the FLRW metric in the $\Lambda$CDM model, and (b) when the gravitational wave propagates through an inhomogeneous spacetime, described by our model within the Buchert framework. The luminosity distance of compact binaries from which gravitational waves have been detected has a massive variation from the order of 50 Mpc to 5000 Mpc \cite{catalog-1, catalog-2}. The length scales of the cosmic voids, which are vast spaces between large-scale structures in the Universe containing no or a negligible number of galaxies, are typically 10 to 100 Mpc \cite{Baushev}. So, it is pretty clear that for the typical sources responsible for detection of gravitational wave events by aLIGO and VIRGO, the gravitational waves might have to traverse through one or more of these voids while reaching the earth. Hence, the physical interpretation of our model is that all the cosmic voids in the path of propagation of the gravitational wave constitute the underdense region and the rest of the regions rich in galaxies and other stellar matters constitute the overdense region.
 
\par Then, according to the definitions of average quantities as given in the equations (\ref{1.2}) and (\ref{1.3}), the average volume scale-factor $a_{\mathcal{D}} $ and the average Hubble-parameter for our two-partitioned model are given by respectively :
\begin{equation} \label{2.1}
a_{\mathcal{D}} = \Big(  \frac{a^{3}_{u}   +  a^{3}_{o}}{ a^{3}_{u, 0}   +  a^{3}_{o, 0} }  \Big)^{1/3}  \, , 
\end{equation}
and, 
\begin{equation} \label{2.2}
H_{\mathcal{D}} = H_{u} \frac{a^{3}_{u}}{ a^{3}_{u}   +  a^{3}_{o} } + H_{o} \frac{a^{3}_{o}}{ a^{3}_{u}   +       a^{3}_{o} } \, , 
\end{equation}
where the $a_{u} $ and $a_{o}$ denote the volume scale-factors of the underdense region and overdense region respectively. The suffix `0' stand for the present time, {\it i.e.}, $a_{u, 0} $ and $a_{o, 0}$ denote the present values of these scale-factors.  $H_{u}$ and $H_{o}$ denote respectively the Hubble-parameters of the underdense and overdense regions.

\par In our model, we consider that the overdense region or `Wall' is described by a closed and dust-only FLRW-region and the underdense region or `Void' is an empty (or, having negligible matter-density) FLRW-region. The backreaction term $\mathcal{Q}_{\mathcal{D}}$ for our 2-partitioned model is given by \cite{Wiegand_et_al},
\begin{equation}\label{1.9}
    \mathcal{Q}_{\mathcal{D}} = \mathcal{Q}_o + \mathcal{Q}_u + 6f_o(1 - f_o)(H_o - H_u)^2,
\end{equation}
where $f_o$ denotes the volume fraction of the overdense region. For our model, we have taken the sub-domains to be described by FLRW-regions to neglect backreaction on the sub-domains for simplification, i.e., $\mathcal{Q}_o = 0$ and $\mathcal{Q}_u = 0$. This stipulation to FLRW is an approximate assumption governing our toy model (in the more general case the sub-domains may not necessarily be FLRW-regions). From equation (\ref{1.9}), it can be seen that control over global backreaction can be achieved only if the individual backreaction terms are not set to zero.

The time-evolution of the scale-factors $a_{u} $ and $a_{o}$, of the underdense region and overdense region in our model, can be described parametrically in terms of a development-angle $\phi$.  We choose the following ansatz which is a generalization of earlier
 formulations \cite{Koksbang, Koksbang-2, AmnaAli}:
\begin{eqnarray}
t = t_{0} \Big(  \frac{\phi-  \sin\, \phi}{\phi_{0} - \sin\, \phi_{0} }  \Big) \, ,  \label{2.3} \\
a_{o} = \frac{f_o^{1/3}}{2} (1 - \cos\, \phi) \, , \label{2.4} \\
a_{u} = \frac{f_u^{1/3} (\phi_{0} - \sin\, \phi_{0} )}{\pi t_{0}}  t^{\beta}  \, \label{2.5}. 
\end{eqnarray}         
In the above set of equations (\ref{2.3}) - (\ref{2.5}), $f_u$ and $f_o = 1- f_u $ are the fractions of volume of the underdense region and overdense region respectively at $\phi = \pi $.  $\phi_{0}$ is the development-angle parameter at the present time and $\phi_{0}= 3\pi / 2 $~\cite{Koksbang-2}. We  set the present age of the Universe $ t_{0} \approx 13.8 \, Gy $.  $a_{u} $ can be expressed in terms of $ \phi$, by substituting the expression of the time $ t $ from equation (\ref{2.3}) on the RHS of the equation (\ref{2.5}). Here, the parameter $\beta$ is generally chosen between $2/3$ to $ 1$,  to denote
any behaviour ranging from a  matter dominated region, {\it viz}., $ \beta = 2/3$, up to a dark energy dominated region {\it viz}., $ \beta  > 1 $. 
\vspace{0.29cm}   
\section{Modification of redshift and distance calculation in the backreaction framework}\label{Sec3}

As Buchert's backreaction formalism is based on spatially averaged quantities over the concerned domain, it is necessary to relate these spatially averaged quantities with observables in cosmology. \footnote{It may be noted that a distance-redshift relation in an inhomogeneous Universe was derived earlier by considering an approximate metric \cite{Futamase_Sasaki}. In our present analysis, we consider a more advanced approach based on the procedure of averaging.} There are two specific schemes for relating spatial averages with observations. These are the template scheme \cite{Larena, Rosenthal, Paranjape} and the covariant scheme. In the covariant scheme proposed by S. R\"{a}s\"{a}nen \cite{Rasanen-3, Rasanen-4}, the effective redshift $z $ and angular-diameter distance $D_{A} $ are given by the equations 
\begin{eqnarray} 
1+z = \frac{1}{a_{\mathcal{D}}} \, , \label{3.1} \\
H_{\mathcal{D}}\frac{d}{dz} \Big( (1+z)^{2} H_{\mathcal{D}} \frac{d D_{A}}{dz} \Big) = - \frac{4 \pi G}{c^{4}} \langle \rho_{\mathcal{D}} \rangle D_{A}   \, .  \label{3.2} 
\end{eqnarray}  
The effective redshift is defined in terms of the averaged scale-factor $a_{\mathcal{D}}$ of the domain $\mathcal{D}$. The covariant scheme can be applied provided that the spatial averages are calculated on the hypersurfaces of statistical homogeneity and isotropy, and the evolution of structure in the Universe is sufficiently slow in the time-interval of propagation of the wave from source to observer. It has been shown \cite{Koksbang, Koksbang-3} that for a model like ours, the covariant scheme is suitable for describing the relation between the effective redshift and cosmological angular diameter distance. 

In the present context, the domain $\mathcal{D}$ on which we take the averages, encompasses the path of propagation of the gravitational wave from the source to the observer, and could arise from an arbitrary combination of the fractions $fu$ and $fo$. We study the redshift-distance relation for different combinations of these fractions. In fig.\ref{Fig1} we plot the ratio of the angular diameter distance $D_{A}$, given by equation (\ref{3.2}), to the present Hubble-length scale ($D_{H}= cH_{0}^{-1} $) with the redshift for five different combinations of the fractions $fu$ and $fo$ in our model along with the plot of $D_A/D_H$ for the $\Lambda $CDM model.  

\begin{figure}[h]
\includegraphics[width=7.6cm]{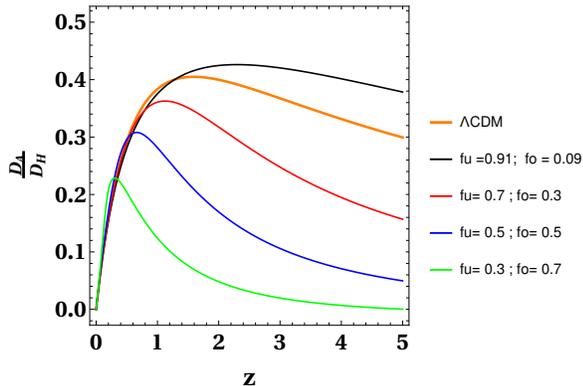}
\caption{Plot of the ratio of angular diameter distance $D_{A}$ to the present Hubble length-scale $D_{H}$ w.r.t. redshift, for the $\Lambda $CDM case with different combinations of the fractions $fu$ and $fo$ for $\beta = 1 $.}\label{Fig1}
\end{figure}

\begin{figure}[h]
\includegraphics[width=8.3cm]{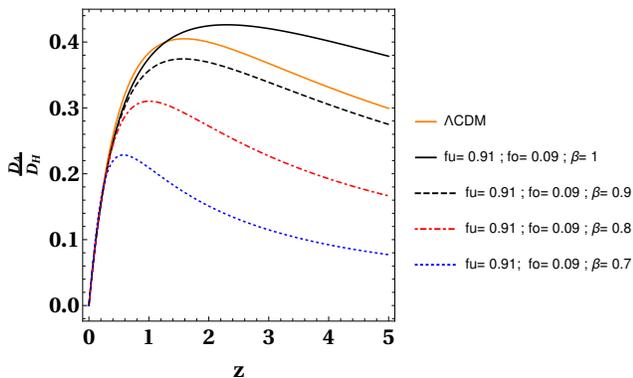}
\caption{Plot of the ratio of angular diameter distance $D_{A}$ to the present Hubble length-scale $D_{H}$ w.r.t. redshift, for the $\Lambda $CDM case with different values of the parameter $\beta $, with the combination of the fractions $fu = 0.91$ and $fo = 0.09$. }\label{Fig2}
\end{figure}

It can be seen from fig.\ref{Fig1}, that up to redshift around 1, almost all the curves are very close to that of the $\Lambda$CDM case, while after a certain redshift, each of these curves start deviating from the $\Lambda $CDM-curve. It can be ascertained that for a certain combination of $fu$ and $fo$, this redshift-distance relation almost coincides with that of the $\Lambda $CDM case. By numerical calculations using \textit{Mathematica}, we find that for the combination $fu = 0.845 $ and $fo = 0.155 $, with $\beta = 1 $, variation of the quantity $D_A/D_H$ is almost identical to that of the $\Lambda $CDM model. 

As expected, departure from the $\Lambda $CDM model increases for greater inhomogeneity, denoted by larger variation from the value of volume fractions $(fu,fo) = (0.845,0.155)$. Next, in fig.\ref{Fig2} we plot the ratio of angular diameter distance $D_{A}$ to the present Hubble length-scale $D_{H}$ w.r.t. redshift by varying the parameter $\beta$ for the combination of fractions $fu = 0.91$ and $fo = 0.09$. We chose this value of combination $(fu,fo)$ as this is the combination, which has been indicated by N-body simulations of structure-formations for a two-partitioned Universe at present time \cite{Wiegand_et_al}. It is seen that conformity with the homogeneous ($\Lambda$CDM) model can be achieved for intermediate values of the void expansion parameter $\beta (\sim 0.92)$, whereas, the limiting cases of matter domination ($\beta \sim 0.7$) and dark energy ($\beta \sim 1$) lead to greater deviation
from the $\Lambda$CDM case.


The change in the observed redshift of a comoving source in the time interval of its observation is known as the redshift-drift. It can be given by $ \delta z  =  \big[\frac{dz}{dt} \big]_{t=t_{0}} \delta t_{0}$, where $[dz/dt]_{t = t_{0}} $ is the time-rate of change of redshift and the suffix $ t = t_{0} $ indicates its value at the observation time which has been taken as the present time $t_{0} $. As the time scale of observation is generally  much lesser than the time scale of cosmological evolution, the expression of the redshift-drift $\delta z $ can be further simplified by applying the Taylor-series expansion, followed by neglecting the higher-order terms, as given below \cite{Koksbang}:
\begin{equation} \label{4.1}
\delta z  = \delta t_{0} \left\lbrace (1+z)H_{0} - H_{e} \right\rbrace = \delta t_{0} (1+z) \left\lbrace \Big[\frac{\partial a }{\partial t }\Big]_{t_{0}} - \Big[\frac{\partial a }{\partial t }\Big]_{t_{e}}   \right\rbrace\, ,   
\end{equation}  
where $H_{0}$ is the Hubble-parameter at the time of observation, which has been taken as the present time  and $H_{e}$ is the Hubble-parameter at the time of emission of the signal.   

\begin{figure}[h]
\includegraphics[width=9cm]{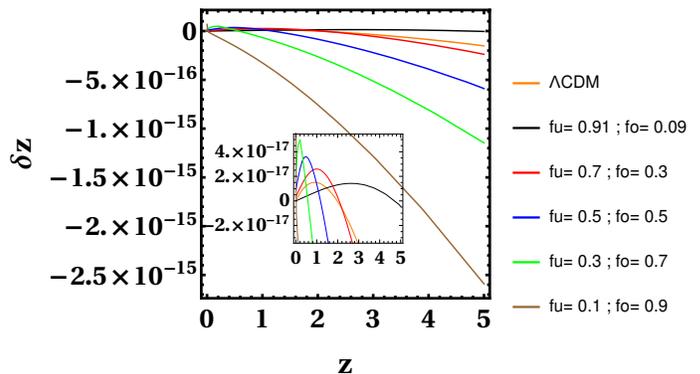}
\caption{Plot of drift of averaged redshift $\delta z $ w.r.t. the redshift $z$, for the $\Lambda$CDM model and for our model 
with different combinations of $fu$ and $fo$ with $\beta = 1$.  The magnified region of the curves, for range of $\delta z $  between $5 \times 10^{-17} $ to $-3 \times 10^{-17} $ is shown in inset in order to ascertain the transition of $\delta z $ from positive to negative value. }\label{Fig3}
\end{figure}   

The redshift-drift  has been discussed in several works \cite{Koksbang, koksnew1, koksnew2, Koksbang-2, heinesen2}. The analysis appears to be the same for the electromagnetic wave and gravitational wave sources, and the redshift-drift is directly proportional to the observation-time $ \delta t_{0} $. However, the redshift drift is minimal for gravitational wave sources; specifically, the binaries of compact objects,  {\it viz}., black holes and neutron stars, which have been observed by aLIGO or VIRGO ($ \delta t_{0} $ is of the order of fractions of seconds to few minutes). Hence, the magnitude of the redshift-drift would be of very low order in these cases. However, for binaries of supermassive black holes, from which the gravitational wave signal is expected to be detectable by the future gravitational wave detectors like LISA, the time of observation $ \delta t_{0} $ can be of much higher order and hence, the redshift-drift's magnitude should increase significantly for these type of binaries, especially at higher redshifts. 

The sign of the redshift-drift indicates whether the Universe is accelerating or decelerating in the time interval of emission and observation of the signal, while its magnitude indicates the extent of the acceleration or deceleration. The expression of redshift-drift given in the equation (\ref{4.1}) implies that for an accelerating Universe between the emission and observation times, the redshift-drift is positive. For a decelerating Universe between those times, it is negative. For the $\Lambda$CDM model, we get the acceleration from a certain time in the late Universe, from which $\Lambda $ starts dominating over the matter till the present. So, for the $\Lambda$CDM model, we can get a positive redshift-drift only after a certain redshift. For an inhomogeneous Universe described by Buchert's backreaction formalism, the drift of the averaged redshift can be effectively described by replacing the Hubble-parameters' values with the averaged Hubble-parameters \cite{Koksbang}. However, as pointed out in Ref. \cite{Koksbang}, the exact calculation of redshift-drift is obtained through a more elaborate process, and the drift of the averaged redshift may not represent the exact redshift-drift correctly.

The main motivation of our present analysis is to study the effect of inhomogeneities on parameters of gravitational wave sources, as discussed in detail in the next section. For the present, we will provide a calculation of the drift of averaged redshift. This is to reemphasize the role of inhomogeneities on the propagation of various signals (electromagnetic and gravitational) in the Universe, as computed through the backreaction formalism. In fig.\ref{Fig3}, we plot the drift of averaged redshift with respect to the redshift for the $\Lambda$CDM model and our model with different combinations of the volume fractions $fu$ and $fo$. By numerical calculations using \textit{Mathematica}, we find that for the combination $fu = 0.768 $ and $fo = 0.232 $, with $\beta = 1 $, variation of the drift of averaged redshift  is almost identical to that of the $\Lambda $CDM model. \footnote{We choose a time-interval of observation $\delta t_{0} =$ 30 seconds, as for the typical binaries of compact objects, from which gravitational wave signals are detectable by aLIGO and VIRGO, the signal is observed within a time-duration of the order of fractions of seconds to minutes at most. }

It may be noted from fig.\ref{Fig3}  that the drifts of averaged redshift are positive only up to a certain redshift for all the curves. The redshift at which the value of the drift of averaged redshift transits from positive to negative is different for each curve. This redshift of transition gradually decreases with an increase in $fo$ or decrease in $fu$.  

\section{Change in  gravitational wave observables in the backreaction framework}\label{Sec4}

The gravitational wave amplitude from a binary of compact objects of masses $m_{1}$ and $m_{2}$, in early inspiral stage where Keplerian approximations are well valid, is given by (for cross($ \times $)-polarization) \cite{maggiore} 
\begin{equation} \label{3.3} 
h_{\times } = \frac{G^{5/3}(1+z)^{5/3}}{D_{L} c^{4}}\frac{m_{1} m_{2}}{(m_{1} + m_{2})^{1/3}}(-4 \omega^{2/3}) \sin \, 2\omega t  \, ,     
\end{equation} 
where $ \omega $ is the observed angular frequency of the binary of compact objects  and $D_{L}$ is the luminosity-distance of the binary from the observer. For the plus($+$)-polarization, the peak of the amplitude remains identical. The amplitude of the gravitational wave at the detector depends on the redshift at which it was generated. For a constant observed frequency, the redshift-dependent part in the gravitational wave amplitude is $(1+z)^{5/3}/D_{L} $. It can also be expressed as $(1+z)^{2/3}/D $, since $D_{L}= (1+z)D $, where $D$ is the cosmological-radial distance of the source. We study the variation of the quantity $ (1+z)^{5/3}/D_{L}$ for the two cases: (i) a homogeneous and isotropic spacetime described by a flat FLRW metric in the $\Lambda$CDM model, and (ii) an inhomogeneous region described by the backreaction formalism, keeping in view that in these two cases the variations of redshift are different. 

\par To depict the change of variation of this quantity $(1+z)^{5/3}/D_{L}$ in the $\Lambda$CDM model of the Universe and in the inhomogeneous Universe described by our two-partitioned model based on the Buchert's backreaction formalism, we plot it w.r.t. redshift $z$, in fig.\ref{Fig4} for these two cases \footnote{ We have kept the upper-limit of the redshift as 5, as accepted theories of cosmology predict the beginning of the structure-formation at the redshift 6 and without a proper structure formation this backreaction formalism can not be applied to model the Universe. For the $\Lambda$CDM model, we have taken the values of $ \Omega_{M} = 0.31$ and $\Omega_{\Lambda} = 0.69 $, with negligible $\Omega_{R}$, as is supported by most of the cosmological observations.} .

For the inhomogeneous Universe, we use a set of different values of the fractions $fo$ and $fu$, as was defined earlier for our two-partitioned model. 

\begin{figure}[h]
\includegraphics[width=8cm]{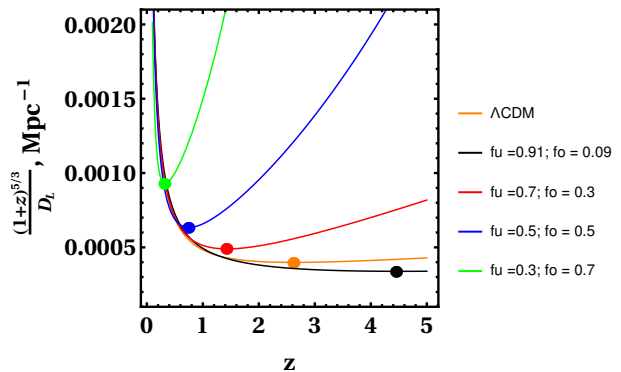}
\caption{Plot of the redshift dependent part of the gravitational wave amplitude for various combinations of the volume fractions $fo$ and $fu$, with $ \beta = 1 $. Position of the minima in curves have been denoted by dots. The shifting of the minima points is
clearly visualized. }\label{Fig4}
\end{figure}
       
\begin{figure}[h]
\includegraphics[width=9cm]{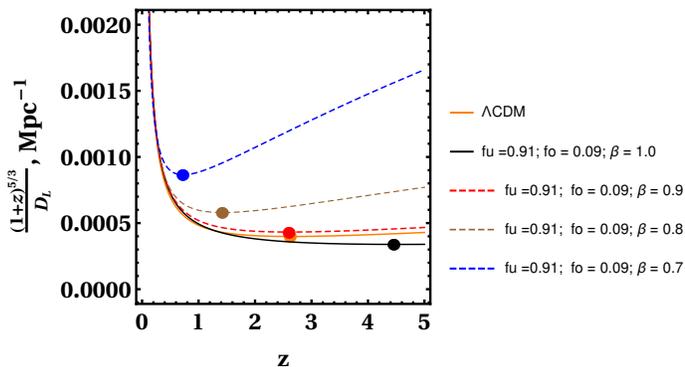}
\caption{Plot of redshift dependent part of the gravitational wave amplitude  w.r.t. redshift $z$, for the $\Lambda$CDM model and for our model for different values of the parameter $\beta$, for the combination of fractions $fu = 0.91$ and $fo = 0.09$. Position of the minima in curves have been denoted by dots. The shifting of the minima points is clearly visualized. }\label{Fig5}
\end{figure}

We see from fig.\ref{Fig4} that as the source-redshift is increased, the deviation in the $(1+z)^{5/3}/D_{L} $ increases among different cases considered here. By numerical calculations using \textit{Mathematica}, we find that for the combination $fu = 0.845 $ and $fo = 0.155 $, with $\beta = 1 $, variation of the quantity $(1+z)^{5/3}/D_{L}$ is almost identical to that of the $\Lambda $CDM model. The deviation of this term for our two-partitioned model compared to that of the $\Lambda$CDM-model increases immensely as we vary the values of $fu$ and $fo$ from the previously mentioned value. We also note that the values of the model parameters, at which the variation of the redshift dependent part of the gravitational wave amplitude with respect to redshift in our model becomes almost identical to the $\Lambda$CDM-case, are different from the case for the drift of averaged redshift.

Now, we study the variation of $ (1+z)^{5/3}/D_{L} $ w.r.t. $z$ for different values of the parameter $\beta$, for the combination of fractions $fu = 0.91$ and $fo = 0.09$. We see from the fig.\ref{Fig5} that as the source-redshift is increased, the deviation in the term $(1+z)^{5/3}/D_{L} $ increases for different curves shown in this graph. Again, as $\beta$ is gradually decreased from the value 1, the deviation in the term $(1+z)^{5/3}/D_{L} $ increases for different curves.
Using \textit{Mathematica}, we find that for the value of the parameter $\beta = 0.92 $, with the combination $fu = 0.91 $ and $fo = 0.09 $, the variation of the quantity $ (1+z)^{5/3}/D_{L} $ in our model is almost identical to that of the $\Lambda$CDM model.
\vspace{0.2cm}\\   
Therefore, it is evident that the observed gravitational wave amplitude is subject to change in our model, in comparison to that of the $\Lambda$CDM model, while the amount of change depends on the effect of inhomogeneities in the domain consisting of the path of propagation of the gravitational wave signified by the model parameters $fu$, $fo$, and $\beta$. 

An interesting consequence of the changing variation of the redshift-dependent part of gravitational wave amplitude   for our model in comparison with the $\Lambda $CDM-model follows from noting the fact \cite{Rosado_et_al} that the quantity $(1+z)^{5/3}/D_{L}  $ has a minimum at $ z_{min}$ given by 
\begin{equation} \label{3.4}
(1+z_{min}) \left[ \frac{d}{dz} ln[D_{L}] \right]_{z= z_{min}} = \frac{5}{3} \,
\end{equation}
We note from the figs.\ref{Fig4} and \ref{Fig5}, that the quantity $ (1+z)^{5/3}/D_{L} $ has different minima  from that of the $\Lambda $CDM-model with variation of the model parameters $fu$, $fo$ and $\beta $. It has been argued in Ref. \cite{Rosado_et_al} that the equation (\ref{3.4}) is independent of the gravitational wave detector, the model of cosmology under consideration, as well as the binary's characteristics, provided its rest-frame frequency is in the inspiral phase. However, our study clearly indicates that although the equation (\ref{3.4}) remains the same, the solution it gives for $z_{min} $ varies in case of our backreaction model. For the $\Lambda$CDM-model the value of this minimum is $z_{min} \approx 2.63 $. We note from fig.\ref{Fig4} that $ z_{min}$  decreases gradually with the increase in $fo$ or decrease in $fu$ for fixed $\beta $. Besides this shifting of the $z_{min} $, the minimum values of the quantity $ (1+z_{min})^{5/3}/D_{L}(z_{min}) $ increases with increase in $fo$ or decrease in $fu$. Again, we note from fig.\ref{Fig5} that for the fixed combination $fu = 0.91$ and $fo = 0.09$, $z_{min} $ decreases and the minimum value of the quantity $ (1+z_{min})^{5/3}/D_{L}(z_{min}) $ increases if we decrease the parameter $\beta $. This shifting can be said to be an effect of the local inhomogeneities that have been considered in our model.  

\section{Conclusions}\label{Sec5}

In this work, we have studied the propagation of gravitational waves from compact binary sources through a background spacetime which due to the presence of structures in the late time Universe, may not be correctly described by the flat FLRW metric at the relevant scales. In order to take into account the effect of inhomogeneities we have employed the Buchert's backreaction formalism \cite{Buchert-1, Buchert-2, Wiegand_et_al} in the framework of which we consider a two-partitioned toy model. Evolution under this model is shown to lead to a modification in the redshift versus distance relation as well as in the drift of the averaged redshift in comparison with the $\Lambda$CDM model, in agreement with similar earlier results \cite{Koksbang}within the backreaction framework.

Our analysis exhibits a substantial deviation in the variation of the redshift dependent part of the amplitude of gravitational waves generated from binaries of compact objects in their early inspiral stages, in the two cases, {\it  viz.}, (i) the $ \Lambda$CDM model, and (ii) our model based on Buchert's backreaction formalism. We have investigated the latter model for various combinations of the model parameters such as the volume fractions of overdense and underdense regions, as well as the rate of expansion of the underdense region.  We find that only for very specific combinations of the above parameters, the variation of the redshift dependent part of the amplitude of gravitational wave matches with that of the $\Lambda $CDM case. The deviation of the variation increases with the increasing effect of inhomogeneities quantified by the volume fractions and the void expansion rate. Our results further display an interesting shift in
the minima of the redshift dependent part of the gravitational wave amplitude, which is a clear consequence of the  backreaction from inhomogeneities.

Before concluding, we would like to make the following comments. In the
present era of precision cosmology with several upcoming probes, the
determination of properties of distant sources of gravitational waves will play an important role in several domains of astrophysics and cosmology. The significance of precise determination of gravitational wave observables have spurred many recent works on ascertaining the role of effects such as primordial black hole accretion \cite{bellido,arnab},
and various phase transitions and inflationary scenarios \cite{guo}.
On the other hand, our present work investigates the role of observed structure \cite{Sloan, WiggleZDE} in the Universe on gravitational wave propagation, using the backreaction framework  \cite{Buchert-1, Buchert-2, Wiegand_et_al} without invoking non-standard physics. Through our analysis we have found a clear signature of backreaction in the shift of the gravitational wave amplitude minima in the context of our simplified two-domain model. Our present analysis should inspire further detailed calculations using more sophisticated backreaction scenarios, such as multi domain models, in order to provide more accurate predictions of the quantitative deviation of gravitational wave observables from those obtained in $\Lambda$CDM cosmology.

 \vspace{0.2cm}
{\it Acknowledgements:}
The authors thank Sophie Marie Koksbang for an important comment related to our work. The authors would also like to thank the anonymous referee whose comments have led to significant improvements in the presentation. SSP thanks the Council of Scientific and Industrial Research (CSIR), Govt. of India, for funding through CSIR-JRF-NET fellowship. AS thanks S. N. Bose National Centre for Basic Sciences, Salt lake city, Kolkata-700106, under Department of Science and Technology (DST), Govt. of India, for funding through institute-fellowship.
\end{small}

\end{document}